\documentclass[11pt]{article}
\usepackage{dsfont}
\usepackage{appendix}
\usepackage{curves}
\usepackage{mathrsfs}
\usepackage{pifont}
\usepackage{latexsym,amsthm,amsmath,amssymb,amsfonts,epsfig,graphicx,cite,psfrag}
\usepackage{eepic,color,colordvi,amscd}
\usepackage{tikz}
\usepackage{lineno}      

\newtheorem{theorem}{Theorem}[section]

\newtheorem{lemma}{Lemma}[section]
\newtheorem*{lem37}{Lemma 3.7}
\newtheorem*{lem38}{Lemma 3.8}
\newtheorem*{lem39}{Lemma 3.9}
\newtheorem*{lem310}{Lemma 3.10}
\newtheorem*{lem311}{Lemma 3.11}
\newtheorem*{lem312}{Lemma 3.12}

\newtheorem{conjecture}{Conjecture}

\renewcommand{\baselinestretch}{1.2}

\def\qed{\hfill \rule{4pt}{7pt}}

\def\pf{\noindent {\it Proof. }}

\newtheorem{define}{Defination}[section]
\begin{document}

\title{On the chromatic number of a family of odd hole free graphs\thanks{\scriptsize
Partially supported by NSFC$11931006$}}
 \author{\small  Jialei Song\footnote{Email: 756959624@qq.com} \;\; and\;\; Baogang  Xu\footnote{Email: baogxu@njnu.edu.cn OR baogxu@hotmail.com. Corresponding author}\\\\
 \small Institute of Mathematics, School of Mathematical Sciences\\
\small Nanjing Normal University, 1 Wenyuan Road,  Nanjing, 210023,  China\\
}
\date{}

\maketitle


\begin{abstract}
A hole is an induced cycle of length at least 4, and an odd hole is a hole of odd length. A {\em full house}  is a graph composed by a vertex adjacent to  both ends of an edge in $K_4$.
Let $H$ be the complement of a cycle on 7 vertices. Chudnovsky et al \cite{CRST10} proved that every (odd hole, $K_4$)-free graph is 4-colorable and is 3-colorable if it does not has $H$ as an induced subgraph. In this paper, we use the proving technique of Chudnovsky et al to generalize this conclusion to $($odd hole, full house$)$-free graphs, and prove that for $($odd hole, full house$)$-free graph $G$, $\chi(G)\le \omega(G)+1$, and the equality holds if and only if $\omega(G)=3$ and $G$ has $H$ as an induced subgraph.\end{abstract}

\begin{flushleft}
\end{flushleft}

\section{Introduction}

In this paper, we only consider finite and simple graphs. Let $G$ be a graph, and let $X$ be a subset of the vertex set $V(G)$ of $G$. We use $G[X]$ to denote the subgraph of $G$ induced by $X$. We say that $X$ is a {\em stable set} if $G[X]$ has no edges, say that $X$ is a {\em clique} if $G[X]$ is a complete graph, and say that $X$ is a {\em cutset} if $G-X$ (the subgraph induced by $V(G)\setminus X$) has more components than $G$. The {\em clique number} of $G$, denote by $\omega(G)$, is the maximum size of a clique of $G$.


Let $u$ be a vertex of $G$. We use $N_X(u)$ to denote the set of neighbors of $u$ in $X$ (and simply write $N(u)$ if $X=V(G)$). We say that $u$ is {\em complete} to $X$ if $N_X(u)=X$, and say that $u$ is {\em anticomplete} to $X$ if $N_X(u)=\emptyset$. Let $Y$ be a subset of $V(G)\setminus X$. We say that $Y$ is {\em complete} to $X$ if each vertex of $Y$ is complete to $X$, and say that $Y$ is {\em anticomplete} to $X$ if each vertex of $Y$ is anticomplete to $X$.

Let $k$ be a positive integer. We use $[k]$ to denote the set $\{1, 2, \ldots, k\}$.

A $k$-coloring of $G$ is a partition of $V(G)$ into $k$ stable sets, of which each is called a {\em color class}. We say that $G$ is $k$-colorable if $G$ admits a $k$-coloring. The {\em chromatic number}, denoted by $\chi(G)$, of $G$ is defined to be the  minimum integer $k$ such that $G$ is $k$-colorable.
It is certain that $\chi(G)\ge \omega(G)$.

A {\em hole} in  $G$ is an induced cycle of length at least 4, and an {\em odd} (resp. {\em even}) hole is a hole of  odd (resp. even) length.  An {\em antihole} is the complement of a hole, and an {\em odd} (resp. {\em even}) antihole is an antihole with odd (resp. even) number of vertices. Similarly, we define {\em even} (resp. {\em odd} paths.

For a given graph $H$, we say that $G$ induces $H$ if $G$ has an induced subgraph isomorphic to $H$, and say that $G$ is $H$-{\em free} if it does not induce $H$. Let ${\cal H}$ be a family of graphs, we say that $G$ is ${\cal H}$-{\em free} if $G$ is $H$-free for each member $H\in {\cal H}$.

A graph is said to be {\em perfect} if $\chi(H)=\omega(H)$ for each of its induced subgraphs $H$. In 2006, Chudnovsky, Robertson, Seymour and Thomas \cite{CRST2006} confirmed the Strong Perfect Graph Conjecture of  Berge \cite{CB1960, CB1961}, and proved that a graph is perfect if and only if it is (odd hole, odd antihole)-free. Confirming a conjecture of Gy\'{a}rf\'{a}s \cite{gyarfas1}, Scott and Seymour \cite{SS2015} proved that $\chi(G)\le {2^{2^{\omega(G)+2}}\over 48(\omega(G)+2)}$ for every odd hole free graph $G$.

We say that $G$ is 2-{\em divisible} if for each induced subgraph $H$ of $G$, $V(H)$ is either a stable set, or can be partitioned into two sets $X$ and $Y$ such that $\max\{\omega(H[X]), \omega(H[B])\}<\omega(H)$ (i.e., both $X$ and $Y$ intersect every maximum clique of $H$).
It is easy to see that $\chi(G)\le 2^{\omega(G)-1}$ for each 2-divisible graph $G$. Ho\'{a}ng and McDiarmid \cite{HM02} introduced this concept, and proved that $($odd hole, $K_{1, 3})$-free graphs are 2-divisible. They also posed the following conjecture.

\renewcommand{\baselinestretch}{1}
\begin{conjecture}\label{H-M-Conj}
A graph is $2$-divisible if and only if it is odd hole free.
\end{conjecture}\renewcommand{\baselinestretch}{1.2}

In \cite{H18}), Ho\'{a}ng introduced and studied another nice property of graphs--the {\em perfect divisibility}.  A graph $G$ is called \textit{perfectly divisible} if for each induced subgraph $H$, $V(H)$ can be partitioned into two subsets $A$ and $B$ such that $H[A]$ is perfect and $\omega(H[B])<\omega(H)$. There exist perfectly divisible graphs which are not 2-divisible (such as odd holes). If $G$ is perfectly divisible, then $\chi(G)\leq {\omega(G)+1\choose 2}$.
In \cite{SS2020}, Scott and Seymour mentioned a conjecture of Ho\'{a}ng.

\renewcommand{\baselinestretch}{1}
\begin{conjecture}\label{H--Conj}
If a graph $G$ is odd hole free, then $V(G)$ can be partitioned into $\omega(G)$ subsets of which each induces a perfect graph.
\end{conjecture}\renewcommand{\baselinestretch}{1.2}

Chudnovsky, Robertson, Seymour and Thomas \cite{CRST10} confirmed Conjectures~\ref{H-M-Conj} and \ref{H--Conj} on $K_4$-free graphs. They proved the following

\renewcommand{\baselinestretch}{1}
\begin{theorem}\label{th-CRST10}
Every $($odd hole, $K_4)$-free graph is $4$-colorable.
\end{theorem}\renewcommand{\baselinestretch}{1.2}

For more recent results on Conjectures~\ref{H-M-Conj} and \ref{H--Conj}, interested readers are referred to \cite{CS19, H18}.

A {\em full house}  is a graph composed by a vertex adjacent to both ends of an edge in $K_4$.
It is certain that $K_4$-free graph must be full house free. In this paper, we study (odd hole, full house)-free graphs, and prove the following

\renewcommand{\baselinestretch}{1}
\begin{theorem}\label{main}
Let $G$ be an $($odd hole, full house$)$-free graph. Then, $\chi(G)\le \omega(G)+1$, and the equality holds if and only if  $\omega(G) = 3$ and $G$ induces an odd antihole on seven vertices.
\end{theorem}\renewcommand{\baselinestretch}{1.2}

As a direct consequence, one sees that both Conjectures~\ref{H-M-Conj} and \ref{H--Conj} hold on full house free graphs.

To prove Theorem~\ref{main}, we follow the idea and the structure of proofs of  Chudnovsky, Robertson, Seymour and Thomas from \cite{CRST10}. We will prove a structural theorem claiming that if $G$ is $($odd hole, full house$)$-free, then either $G$ has a particular cutset, or belongs to two kinds of specific families of graphs.

In Section 2, we present some useful definitions, state our structural theorem, and then prove Theorem~\ref{main}. The proof of the structural theorem will be given in Section 3.

\section{Harmonious cutset, heptagram-type graphs, and $T_{11}$-type graphs}

Let $A$ and $B$ be two subsets of $V(G)$, let $a\in A$ and $b\in B$. An $ab$-path is a path connecting $a$ and $b$, and an $ab$-path $P$ is called an $AB$-{\em path} if $V(P)\cap A=\{a\}$ and $V(P)\cap B=\{b\}$.


A cutset $X$ is said to be {\em harmonious} if $X$ can be partitioned into disjoint sets $X_1, X_2, \dots, X_k$ such that

\renewcommand{\baselinestretch}{1}
\begin{itemize}
\item  \noindent  if $k\ge 3$, then $X_1, \dots, X_k$ are pairwisely complete to each other, and 	
\item  \noindent  for all $i, j\in [k]$, if $P$ is an induced $X_iX_j$-path, then $P$ is even if $i=j$ and odd otherwise.
\end{itemize}\renewcommand{\baselinestretch}{1.2}

From the second condition, one sees that each $X_i$ is a stable set. Let $X$ be a harmonious cutset of  $G$. Suppose that $V(G)\setminus X$ is partitioned into two nonempty sets $C_1$ and $C_2$, and let $G_i=G[C_i\cup X]$ for $i\in [2]$. Following several lemmas, due to Chudnovsky et al \cite{CRST10}, are all  about harmonious cutsets.

\medskip


\renewcommand{\baselinestretch}{1}
\begin{lemma}\label{no odd hole} {\em (\cite[Lemma 2.1]{CRST10})}
If $G_i$ is odd hole free for each $i\in [2]$, then $G$ is odd hole free.
\end{lemma}

\renewcommand{\baselinestretch}{1}
\begin{lemma}\label{harmonious set}{\em (\cite[Lemma 2.3]{CRST10})}
Let $G$ be an odd hole free graph, and let $X$ be a cutset of $G$.	Let $X_1, X_2, \dots, X_k$ be a partition of $X$ into stable sets, such that if $k\ge 3$, then the sets $X_1, X_2, \dots, X_k$ are pairwisely complete. Suppose that for all nonadjacent $a, b\in X$, there exists an induced $ab$-path $P$, with interior in $V(G)\setminus X$, such that $P$ is even if some $X_i$ contains both $a$ and $b$, and odd otherwise. Then $G$ has a harmonious cutset.
\end{lemma}\renewcommand{\baselinestretch}{1.2}

In \cite{CRST10}, Chudnovsky {\em et al} also proved that if $G_i$ is $4$-colorable for $i\in [2]$  then $G$ is $4$-colorable. With the same technique as that of Chudnovsky {\em et al}, we extend the conclusion to $k$-colorability.

\renewcommand{\baselinestretch}{1}
\begin{lemma}\label{k-colourable}
If $G_i$ is $k$-colorable for $i=1, 2$, then $G$ is $k$-colorable.
\end{lemma}\renewcommand{\baselinestretch}{1.2}
\pf  Let $X_1, X_2, \ldots, X_t$ be as the definition of a harmonious cutset. By the hypothesis, both $G_1$ and $G_2$ are $k$-colourable. The lemma holds trivially when $k=1$ or $2$. Then we may suppose $k\ge 3$, and so $t\le k$ by the definition of a harmonious cutset.

For $i\in [2]$, let $\phi$ be a $k$-colouring of $G_i$ using colours $[k]$. A vertex $v\in X_j$ is called a $\phi$-{\em compliant} if $\phi(v)=j$. We will show that
\begin{equation}\label{eqa-k-colorability-1}
\mbox{$G_i$ has a $k$-colouring $c_i$ such that every vertex of $X$ is a $c_i$-compliant}.
\end{equation}

We only need to verify the case that $i=1$ by symmetry. Let $c$ be a $k$-colouring of $G_1$ that maximises the number of $c$-compliant vertices. We will show that $c$ is a $k$-colouring of $G_1$ satisfying the requirement of (\ref{eqa-k-colorability-1}).

Suppose to its contrary, let $v\in X$ be a vertex that is not $c$-compliant, say $v\in X_{j_1}$ and $c(v)=j_2$, where $j_1\ne j_2$. Let $H$ be the component containing $v$ of the subgraph of $G_1$ induced by the vertices coloured $j_1$ or $j_2$. We claim that no vertex of $V(H)\cap X$ is a $c$-compliant.
For otherwise, let $u\in V(H)\cap X$ be a $c$-compliant, and let $P$ be an induced $uv$-path of $H$. Since $u$ is a $c$-compliant by our assumption, $c(u)=j_r$ if and only if $u\in X_{j_r}$ for some $r\in [2]$. From the definition of $H$, $c(u)=c(v)$ if and only if $P$ has even length. Note that $X$ is a harmonious cutset,  $P$ has even length if and only if $\{u, v\}\subseteq X_h$ for some $h\in [t]$.  If $P$ is even, then $\{u, v\}\subseteq X_{j_1}$ as $X$ is a harmonious cutset, and so $c(u)=j_1\ne j_2=c(v)$, a contradiction. If $P$ is odd, then $v\in X_{j_1}$ and $u\in X_{j_2}$ as $X$ is a harmonious cutset, and so $c(u)=j_2=c(v)$, a contradiction as well.  Therefore, no vertex of $V(H)\cap X$ can be $c$-compliant.

Let $c'$ be the colouring obtained from $c$ by swapping the colours $j_1$ and $j_2$ for every vertex of $H$. Then $v$ is $c'$-compliant and it follows that more vertices in $X$ are $c'$-compliants than that of $c$-compliants, contrary to the choice of $c$.
This proves (\ref{eqa-k-colorability-1}).

Now the colorings $c_1$ and $c_2$ can contribute a $k$-colouring of $G$. \qed

\medskip

The following definitions of $T_{11}$-{\em type} graphs and {\em heptagram}-{\em type} graphs are totally the same as that of \cite{CRST10}. We present them here for completeness.
Let $k$ and $l$ be two positive integers and $k<l$. Throughout the paper, by index arithmetic modulo $l$, we always mean that $l+k\equiv k$.

\begin{define}\label{def-t11}
A graph is of $T_{11}$-type if there is a partition of $V(G)$ into eleven nonempty stable sets $W_1, \dots, W_{11}$, such that for $i\in [11]$, $W_i$ is anticomplete to $W_{i+1}\cup W_{i+2}$ and complete to $W_{i+3}\cup W_{i+4}\cup W_{i+5}$,  with index arithmetic modulo 11.
\end{define}

Let $X$ and $Y$ be two disjoint subsets of $V(G)$. We say that $X$ and $Y$ are {\em linked} if
every vertex of $X$ has a neighbor in $Y$, and every vertex of $Y$ has a neighbor in $X$.

\begin{define}\label{def-hepta-type}
A graph $G$ is of {\em heptagram}-{\em type} if there is a partition of $V(G)$ into fourteen stable subsets $W_1, W_2, \dots, W_7$ and $Y_1, Y_2, \dots, Y_7$, where $W_1, W_2, \dots, W_7$ are nonempty but $Y_1, Y_2, \dots, Y_7$ may be empty, satisfying the following (with index arithmetic modulo 7).

\renewcommand{\baselinestretch}{1}
\begin{itemize}
\item [$1.$] For $i\in[7]$, $W_i$ is anticomplete to $W_{i+3}$.
	
\item [$2.$] For $2\le i\le 7$, $W_i$ is complete to $W_{i+2}$, and the subsets $W_1$ and $W_3$ are linked.
	
\item [$3.$] For $i\in \{3, 4, 6, 7\}$, $W_i$ is complete to $W_{i+1}$; for $i\in\{1, 2, 5\}$, $W_i$ and $W_{i+1}$ are linked.
	
\item [$4.$] If $v_i\in W_i$ for $i\in [3]$, and $v_2$ is complete to $\{v_1, v_3\}$, then $v_1$ is adjacent to $v_3$.
	
\item [$5.$] If $v_i\in W_i$ for $i\in [3]$, and $v_2$ is anticomplete to $\{v_1, v_3\}$, then $v_1$ is nonadjacent to $v_3$.
	
\item [$6.$] For $i\in [7]$, every vertex in $Y_i$ has a neighbor in each of $W_i, W_{i+3}$ and $W_{i-3}$, and has no neighbor in $W_{i+1}\cup W_{i+2}\cup W_{i-1}\cup W_{i-2}$.
	
\item [$7.$] For $i\in [7]$ and each $y\in Y_i$, let $N_j$ be the set of neighbors of $y$ in $W_j$ for $j\in \{i, i+3, i-3\}$. Then $N_{i+3}$ is complete to $N_{i-3}$, and $N_{i+3}$ is anticomplete to $W_{i-3}\setminus N_{i-3}$, and $N_{i-3}$ is anticomplete to $W_{i+3}\setminus N_{i+3}$, and $N_i$ is complete to $W_{i+1}\cup W_{i+2}\cup W_{i-1}\cup W_{i-2}$.

\item [$8.$] For $i\in [7]$, $Y_i$ is complete to $Y_{i+1}$ and anticomplete to $Y_{i+2}\cup Y_{i+3}$.
	
\item [$9.$] For $i\in [7]$, if $Y_i$ is not complete to $W_{i-3}\cup W_{i+3}$, then $W_{i-3}\cup W_{i+3}$ is complete to $W_{i-2}\cup W_{i+2}$, and $Y_{i-1}\cup Y_{i+1}\cup Y_{i-3}\cup Y_{i+3}$ is empty.
	
\item [$10.$] For $i\in [7]$, at least one of $Y_i, Y_{i+1}$ and $Y_{i+2}$ is empty.
\end{itemize}\renewcommand{\baselinestretch}{1.2}
\end{define}

Now, we can state our structural theorem.

\renewcommand{\baselinestretch}{1}
\begin{theorem}\label{main'}
Let $G$ be an  $($odd hole, full house$)$-free graph which has no harmonious cutset and induces an antihole on seven vertices. Then, $G$ is either of heptagram-type or of $T_{11}$-type.
\end{theorem}\renewcommand{\baselinestretch}{1.2}

Theorem~\ref{main'} will be proved in latter section. As a direct consequence of Theorem~\ref{main'}, we can now prove Theorem~\ref{main}.

\medskip

\noindent{\em Proof of Theorem}~\ref{main}.
Let $G$ be an (odd hole, full house)-free graph. By Theorem~\ref{th-CRST10}, we need only to prove that $\chi(G)=\omega(G)$ if $\omega(G)\ge 4$. By Lemma~\ref{k-colourable}, we may  suppose that $G$ has no harmonious cutsets.  If $G$ does not induce an antihole on seven vertices, then $G$ is perfect. Otherwise, $G$ will be either of heptagram-type or of $T_{11}$-type by Theorem~\ref{main'}, which is $4$-colourable with $\omega(G)=3$.
\qed

\section{Proof of Theorem~\ref{main'}}

This section is devoted to proving Theorem~\ref{main'}. The whole proof, similar to that of \cite{CRST10}, is divided into 4 subsections, respectively considering $T_{11}$-type subgraphs, Heptagrams, $Y$-vertices, and $V$-vertices etc.

\subsection{$T_{11}$-type graphs}

Firstly, we discuss the $T_{11}$-type configuration contained in (odd hole, full house)-free graphs, and prove that if an $($odd hole, full house$)$-free graph has no harmonious cutset and contains an induced $T_{11}$-type  subgraph, then the graph itself is a $T_{11}$-type graph.

Suppose that $X_1, X_2, \ldots, X_n$ are disjoint subsets of $V(G)$. A path of form $X_{i_1}-\cdots-X_{i_h}$ is an induced path $x_{i_1}x_{i_2}\ldots x_{i_h}$, where $x_{i_j}\in X_{i_j}$ for all $1\le i_1,i_2,\cdots,i_h\le n$ (we denote $x_i$ instead of $X_i$ if $X_i$ is a singleton). We use analogous terminology for holes.

\renewcommand{\baselinestretch}{1}
\begin{lemma}\label{T11}
Let $G$ be an $($odd hole, full house$)$-free graph which has no harmonious cutset.
If $G$ induces an $T_{11}$-type  subgraph, then $G$ itself is $T_{11}$-type.
\end{lemma}\renewcommand{\baselinestretch}{1.2}
\pf Let $F$ be an induced $T_{11}$-type  subgraph of $G$ with maximum number of vertices. As in its definition, suppose that $V(F)$ is partitioned into stable sets $W_1, W_2, \ldots, W_{11}$, let $W=\cup_{i=1}^{11} W_i$, and suppose that $W\ne V(G)$.

Let $X$ be a subset of $V(G)\setminus W$ such that $G[X]$ is connected, and let  $a$ and $b$ be two nonadjacent vertices of $W$ of which each has neighbors in $X$. We claim that
\begin{equation}\label{eqa-T11-1}
\{a, b\}\subseteq W_i \mbox{ for some } i\in [11].
\end{equation}

If it is not the case, we choose $X$ to be minimal, and let $a, b\in W$ be a pair of vertices violating  (\ref{eqa-T11-1}). By symmetry, we may assume  that $a\in W_1$ and $b\in W_2\cup W_3$. The minimality of $X$ implies that there is a positive integer $k$ and an induced path $ax_1\ldots x_kb$ with $X=\{x_1, \ldots, x_k\}$. First we show that
\begin{equation}\label{k>1}
k>1.
\end{equation}

For otherwise, suppose that $X=\{v\}$. Let $N=N_W(v)$. An induced path $P$ is called a $v$-{\em path} if $P$ has both ends in $N$ and interior vertices in $W\setminus N$. As $G$ is odd hole-free, a $v$-path is of length one if it is odd. We suppose that $w_i\in W_i$ for each $i\in [11]$, if necessary.

Suppose first that $b\in W_2$. Since there is no $v$-path of the form $a-W_4-W_{10}-b$, we may suppose by symmetry that $W_4\subseteq N$.
If $v$ has  a neighbor, say $u$, in $W_7\cup W_8\cup W_9$, then $\{a, b, w_4, u, v\}$ induces a full house, and so
$$(W_7\cup W_8\cup W_9)\cap N=\emptyset.$$

Since there is no $v$-path of the form $b-(W_5\cup W_6)-(W_1\cup W_{11})-w_4$, we have that $W_5\cup W_6\subseteq N$ or $W_1\cup W_{11}\subseteq N$.
If $W_5\cup W_6\subseteq N$, then $W_2\subseteq N$ as there is no $v$-path of the form $w_5-W_2-W_7-w_4$, and $W_3\subseteq N$ as there is no $v$-path of the form $w_5-W_8-W_3-w_6$, and $W_1\subseteq N$ as there is no $v$-path of the form $w_5-W_1-W_7-w_3$. Since for each vertex  $u$ in $W_{10}\cup W_{11}$, $\{u, w_3, w_4, w_6, v\}$ cannot be a full house, $(W_{10}\cup W_{11})\cap N=\emptyset$. By putting $v$ into $W_9$, we get a $T_{11}$-type subgraph with one more vertex than $F$. This contradiction shows that $W_5\cup W_6\not\subseteq N$, and so
$$W_1\cup W_{11}\subseteq N$$
which implies that $W_3\subseteq N$ as there is no $v$-path of the form $a-W_9-W_3-w_{11}$.

Since for each vertex $u$ in $W_1$, $\{w_3, w_6, w_{11}, u, v\}$ cannot induce a full house, we have $W_6\cap N=\emptyset$.
As there is no $v$-path of the form $a-W_5-W_{10}-W_3$, we may suppose by symmetry that $W_5\subseteq N$. Consequently, $W_2\subseteq N$ as there is no $v$-path of the form $W_5-W_2-W_7-W_4$, and $W_{10}\cap N=\emptyset$ as $\{w_2, w_5, w_{10}, w_4, v\}$ cannot induce a full house. By putting  $v$ into $W_8$, we get another $T_{11}$-type subgraph with one more vertex than $F$. This contradiction shows that $b\notin W_2$. Therefore,
$$b\in W_3,$$
and by symmetry, we may suppose that $N$ is disjoint from one of $W_i$ and $W_{i+1}$ for each $i\in [11]$. But now, we have that $N\cap (W_{11}\cup W_2\cup W_4)=\emptyset$, and a $v$-path of the form $a-W_4-W_{11}-b$ appears. This contradiction proves (\ref{k>1}).


\medskip

By the minimality of $X$, one of $W_i, W_{i+1}$ and $W_{i+2}$ is anticomplete to $X$. In particular, $W_j$ is anticomplete to $X$ for some $j\in \{6, 7, 8\}$. Since  $w_jax_1x_2\ldots x_kbw_j$ cannot be an odd hole, we have that $k$ is odd.

If $b\in W_2$, since $ax_1\ldots x_kbw_{10}w_4a$ cannot be an odd hole, we may assume by symmetry that $N_X(w_4)\ne \emptyset$ which implies that $N_X(w_4)=\{x_1\}$ by the minimality of $X$ and by the fact that $k$ is odd, then there exists an $h\in \{5, 6\}$ such that $W_h$ is anticomplete to $X$ (since one of $W_4, W_5$ and $W_6$ is anticomplete to $X$), and so an odd hole $w_4x_1\ldots x_kbw_hw_{11}w_4$ appears, a contradiction.

Therefore, $b\in W_3$, and by symmetry, we may assume that for each $i\in [11]$, one of $W_i$ and $W_{i+1}$ is anticomplete to $X$. In particular, $W_2\cup W_4\cup W_{11}$ is anticomplete to $X$. Now, $ax_1\ldots x_kbw_{11}w_4a$ is an odd hole. This contradiction proves (\ref{eqa-T11-1}).

\medskip

Now we turn to prove the lemma. Choose $C\subseteq V(G)\setminus W$ such that $G[C]$ is connected. For each $i$, let $N_{W_i}(C)$ be the set of vertices of $W_i$ which have neighbors in $C$. Let $I\subseteq [11]$ be a maximum set such that $N_{W_i}(C)\ne \emptyset$ for each $i\in I$. By (\ref{eqa-T11-1}), $|I|\le 3$ as $11/3<4$, and $N_{W_i}(C)$ is complete to $N_{W_j}(C)$ for all distinct pairs $\{i, j\}\subseteq I$. Let $W'=\cup_{i\in I} N_C(W_i)$. Then $W'$ is a cutset of $G$ such that for each distinct pair $a, b\in W'$, each induced $ab$-path with interior in $V(G)\setminus W'$ is even if and only if $\{a, b\}\subseteq N_{W_i}(C)$ for certain $i\in I$ (since $N(C)\cap W_4=\emptyset$ or $N(C)\cap W_5=\emptyset$ by (2), let $a,b\in N(C)\cap W_1$, we know that there exists an induced even $ab$-path as $a$ and $b$ have a common neighbor in $W_j$ for $j=4$ or $5$, and so the argument holds).
By Lemma~\ref{harmonious set}, $G$ admits an harmonious cutset. This completes the proof of Lemma~\ref{T11}.  \qed

\medskip

In the following text, we always suppose that $G$ is an (odd hole, full house)-free graph which is not a $T_{11}$-type graph,  contains no harmonious cutset, and induces an odd antihole  on seven vertices. Since $G$ is full house free, it is certain that $G$ induces no odd antihole on at least nine vertices.  We will show that $G$ is a heptagram-type graph.

\subsection{Heptagrams}

Following the proof technique of \cite{CRST10}, we first discuss the local structure around an odd antihole on on seven vertices. Let $F$ be an odd antihole on seven vertices. Then, the vertices of $F$ can be labelled as  $w_1, w_2, \ldots, w_7$ in such way that $w_i$ is adjacent to $w_j$ if and only if $|i-j|\in \{1, 2\}$ (here and later all index summations are modulo 7).

\begin{define}\label{def-hepta}
A \textit{heptagram} $W=(W_1, W_2, \ldots, W_7)$ in $G$ is defined as follows (see \cite{CRST10}).

\renewcommand{\baselinestretch}{0.8}
\begin{itemize}
	\item [$1.$] $W_1, W_2, \ldots, W_7$ are disjoint, nonempty stable sets,
	
	\item [$2.$] for $i\in [7]$, $W_i$ is anticomplete to $W_{i+3} \cup W_{i+4}$,
	
	\item [$3.$] for $i\in [7]$, $W_i, W_{i+1}$ and $W_{i+2}$ are pairwise linked,
	
	\item [$4.$] if $u\in W_{i-1}$, $v\in W_i$, $w\in W_{i+1}$, and $v$ is complete to $\{u, w\}$, then $uw\in E(G)$,
	
	\item [$5.$] if $u\in W_{i-1}$, $v\in W_i$, $w\in W_{i+1}$, and $v$ is anticomplete to $\{u, w\}$, then $uw\not\in E(G)$, and
	
	\item [$6.$] if $u\in W_{i-1}$, $v\in W_i$, $w\in W_{i+1}$, $x\in W_{i+2}$ and $uw,vx\in E(G)$, then either $u$ is adjacent to $v$ or  $w$ is adjacent to $x$.
\end{itemize}
\end{define}

\renewcommand{\baselinestretch}{1.2}

The following three lemmas are from  \cite{CRST10}, which are all on the structure of heptagrams.
\renewcommand{\baselinestretch}{1}
\begin{lemma}\label{i&i+2}{\em (\cite[Lemma 5.1]{CRST10})}
	Let $(W_1, W_2, \ldots, W_7)$ be a heptagram in $G$. For $i\in [7]$, if $W_i$ is complete to $W_{i+1}$, then $W_i$ is complete to $W_{i+2}$ and $W_{i-1}$ is complete to $W_{i+1}$.
\end{lemma}

\begin{lemma}\label{i&i+1}{\em (\cite[Lemma 5.2]{CRST10})}
	Let $(W_1, W_2, \ldots, W_7)$ be a heptagram in $G$. For $i\in [7]$ either $W_i$ is complete to $W_{i+1}$ or $W_{i+2}$ is complete to $W_{i+3}$.
\end{lemma}

\begin{lemma}\label{common path}{\em (\cite[Lemma 5.3]{CRST10})}
	Let $(W_1, W_2, \ldots, W_7)$ be a heptagram in $G$. Then there exists a $t\in [7]$ such that $W_{j-1}$ is complete to $W_{j+1}$ for all $j\in [7]\setminus \{t\}$, and $W_j$ is complete to $W_{j+1}$ for all $j\in \{t-3, t-2, t+1, t+2\}$. Consequently, for all $i\in [7]$, if $u\in W_{i-2}$ and $v\in W_{i+2}$, then
	
	\begin{itemize}
		\item \noindent $u, v$ have  common neighbors in $W_{i-3}$, in $W_i$, and in $W_{i+3}$, and
		\item \noindent there is a path of the form $u-W_{i-1}-W_{i+1}-v$.
	\end{itemize}

\end{lemma} \renewcommand{\baselinestretch}{1.2}


From now on to the end of this section, we always suppose that $(W_1, W_2, \ldots, W_7)$ is a heptagram in $G$, and still use $W$ to denote $W_1\cup W_2\cup\cdots\cup W_7$. As that in \cite{CRST10}, we will define and study the properties of vertices associated to $(W_1, W_2, \ldots, W_7)$.

\subsection{$Y$-vertices}


In this subsection, we study the properties of  $Y$-vertices defined  below (see \cite{CRST10}).

\begin{define}\label{def-y-vertex}
A vertex  $y\in V(G)\setminus W$ is called a \textit{Y-vertex} or a \textit{Y-vertex of type}-$t$ if the following hold, where $N_i=N_{W_i}(y)$ for $i\in [7]$:
\renewcommand{\baselinestretch}{1}
\begin{itemize}
	\item [$1.$] \noindent $N_t, N_{t+3}, N_{t-3}$ are nonempty, and $N_i=\emptyset$ for $i\in \{t-2, t-1, t+1, t+2\}$,
	\item [$2.$] \noindent $N_{t-3}$ is complete to $N_{t+3}$, and $N_{t-3}$ is anticomplete to $W_{t+3}\setminus N_{t+3}$, and $N_{t+3}$ is anticomplete to $W_{t-3}\setminus N_{t-3}$,
	\item [$3.$] \noindent $N_t$ is complete to $W_{t-1}\cup W_{t-2}\cup W_{t+1}\cup  W_{t+2}$.
\end{itemize}
\end{define}

\begin{lemma}\label{Y-ver}
Let $v\in V(G)\setminus W$, and let $N = N_{W}(v)$. Then one of the following holds:
	\begin{itemize}
		\item \noindent $v$ is a $Y$-vertex, or
		\item \noindent $N\cap W_i$ is nonempty for at most two integers $i\in [7]$,	and if there are two such integers, $i$ and $j$ say, then $j\in \{i-2, i-1, i+1, i+2\}$ and $N\cap W_i$ is complete to $N\cap W_j$.
	\end{itemize}

\end{lemma} \renewcommand{\baselinestretch}{1.2}
\pf Without loss of generality, we suppose that $v$ is not a $Y$-vertex. For $i\in [7]$, let  $N_i=N\cap W_i$ and $M_i=W_i\setminus N_i$, and let $I=\{i\in [7] : N_i\ne\emptyset\}$.

\begin{equation}\label{eqa-Y-vertex-1}
|\{i, i+1, i+2, i+3\}\cap I|\le 3 \mbox{ for each } i\in [7].
\end{equation}

Suppose to its contrary that $\{1, 2, 3, 4\}\subseteq I$ by symmetry,  and let $n_i\in N_i$ for $i\in [4]$.
By Lemma~\ref{i&i+1}, we have $W_1$ is complete to $W_2$ or $W_3$ is complete to $W_4$.
We may suppose that $W_1$ is complete to $W_2$ and $n_1n_2\in E(G)$.

Suppose that $G[\{n_1, n_2, n_3\}]$ is a triangle.
Then  $N_{W_7}(n_1)$ is complete to $v$, as otherwise there exists an $n_7\in N_{W_7}(n_1)$ such that $G[\{n_1, n_2, n_3, n_7, v\}]$ is a full house by Definition~\ref{def-hepta}$(4)$.
For any $n'_4\in W_4$, if $G[\{n_2, n_3, n'_4\}]$ is a triangle, then $n'_4v\in E(G)$ as otherwise $G[\{n_1, n_2, n_3, n'_4, v\}]$ is a full house.
But now $G[\{n_2, n_3, n'_4, n_7, v\}]$ is a full house.
This contradiction shows that
\begin{equation}\label{eqa-Y-vertex-2}
G[\{n_1, n_2, n_3\}] \mbox{ cannot be a triangle  and } n_2n_3\notin E(G) \mbox{ by Definition~\ref{def-hepta}$(4)$}.
\end{equation}

By Lemma~\ref{i&i+1}, we see that $W_4$ is complete to $W_5$, and $W_1$ is complete to $W_7$ which implies that $W_2$ is complete to $W_7$ by Definition~\ref{def-hepta}$(4)$.
By Lemma~\ref{i&i+2}, we see that $W_1$ is complete to $W_3$.
Consequently, $N_7=\emptyset$ as otherwise $G[\{n_1, n_2, n_3, n_7, v\}]$ is a full house.
By Lemma~\ref{i&i+2}, $W_3$ is complete to $W_5$. Let $w_7\in W_7$. Since $n_3n_5w_7n_2$ is not a $v$-path and $N_7=\emptyset$, we have that $W_5= N_5$.

By the same argument in showing (\ref{eqa-Y-vertex-2}) and $N_7=\emptyset$, we can show that $G[\{n_3, n_4, n_5\}]$ is not a triangle, and consequently prove that $N_6=\emptyset$.
Particularly, $n_3n_4\notin E(G)$ and $n_2n_4\notin E(G)$ by Definition~\ref{def-hepta}$(5)$.
Let $w_6\in W_6$. Now, $vn_4w_6w_7n_2v$ is a 5-hole $v$-path by Lemma~\ref{common path}, which contributes a contradiction. This proves (\ref{eqa-Y-vertex-1}).

\medskip

Next, we  show progressively that $|I|\le 2$. We start from proving the following
\begin{equation}\label{eqa-Y-vertex-3}
|I|\le 4.
\end{equation}

If (\ref{eqa-Y-vertex-3}) is not true, we may suppose, by (\ref{eqa-Y-vertex-1}) and by symmetry, that $I=\{1, 2, 4, 5, 7\}$.

Choose $n_1\in N_1$. Since $W_{i}$ is linked to $W_{i+1}$ by Definition~\ref{def-hepta}$(3)$, every vertex of $W_2$ has a neighbor in $W_3$. If $n_1$ has a neighbor $n_2\in N_2$ and a neighbor $n_7\in N_7$, let $n_3\in N_{W_3}(n_2)$, then $G[\{n_1, n_2, n_3\}]$ is a triangle by Definition~\ref{def-hepta}$(4)$ which produces a full house on $\{n_1, n_2, n_3, n_7, v\}$. So, we may assume by symmetry that $n_1$ is anticomplete to $N_2$. Now, $n_1$ has a neighbor $w_2\in M_2$ by Definition~\ref{def-hepta}$(3)$.

Since $W_1$ is not complete to $W_2$, $W_3$ is complete to $W_4$ and $W_6$ is complete to $W_7$ by Lemma~\ref{i&i+1},  and consequently $W_3$ is complete to $W_5$ and $W_6$ is complete to $W_1$ by Lemma~\ref{i&i+2}.
Choose $n_2\in N_2$. If $n_2$ has a neighbor $w_1\in M_1$, then a 5-hole $vn_2w_1w_6n_1v$ appears. Therefore, $n_2$ is anticomplete to $M_1$, and so must have a neighbor, say $n'_1$, in $N_1$. Since $n'_1\in N_1$ and $N_{N_2}(n'_1)\ne \emptyset$, by the same argument as above used to $n_1$, we can show that $n'_1$ is anticomplete to $N_7$. Consequently, we have that $n'_1$ has a neighbor $w_7\in M_7$ as $W_1$ and $W_7$ are linked by Definition~\ref{def-hepta}$(3)$, and that $W_2$ is complete to $W_3$ and $W_5$ is complete to $W_6$ by Lemma~\ref{i&i+1}.

By Lemma~\ref{i&i+2}, we have $W_5$ is complete to $W_7$. Let $n_4\in N_4$. If $n_4$ is anticomplete to $N_5$, then $n_4$ has a neighbor $w_5\in M_5$, and so a 5-hole $vn_4w_5w_7n_5v$ appears,  a contradiction. Let $n_5\in N_{N_5}(n_4)$, $w_3\in W_3$ and $w_6\in W_6$.

Since $n'_1n_2\in E(G)$ and $n'_1w_7\in E(G)$, we have that $n_2w_7\in E(G)$ by Definition~\ref{def-hepta}$(4)$.
Let $n_7\in N_7$. Since $n_2n_7\notin E(G)$ would produce a 5-hole $vn_2w_7w_6n_7v$, we see that $n_2n_7\in E(G)$, and so $n_1n_7\in E(G)$ by Definition~\ref{def-hepta}$(5)$ as $n_1n_2\not\in E(G)$.

Consequently, we have  $n_7w_2\in E(G)$ by Definition~\ref{def-hepta}$(4)$, $n'_1w_2\in E(G)$ and $n_1w_7\in E(G)$ by Definition~\ref{def-hepta}$(5)$, and $w_7w_2\in E(G)$ by Definition~\ref{def-hepta}$(4)$. But now,
$G[\{v, w_3, w_7, n_7, n_4, n'_1, n_1, n_5,$ $ n_2, w_2, w_6\}]$ is a $T_{11}$.
This proves (\ref{eqa-Y-vertex-3}).

\medskip

After (\ref{eqa-Y-vertex-3}), we show

\begin{equation}\label{eqa-Y-vertex-4}
|I|\le 3.
\end{equation}

If it is not the case, then $|I|=4$. By (\ref{eqa-Y-vertex-1}), we may assume that $1, 4\in I$. If $N_5=N_7=\emptyset$, a 5-hole of the form $v-N_1-W_7-W_5-N_4-v$ appears by Lemma~\ref{common path}. By symmetry, we assume $5\in I$.

We first consider the case that $6\in I$, i.e., $I=\{1, 4, 5, 6\}$. If $N_4$ is not complete to $N_5$, let $n_4\in N_4$ and $n_5\in N_5$ be two  nonadjacent vertices,
by Definition~\ref{def-hepta}$(3)$, there exists a vertex $w_2\in N_{W_2}(n_4)$, and so there exist a vertex $w_7\in W_7$ such that $w_7\in N(w_2)\cap N(n_5)$ by Lemma~\ref{common path},
then a 5-hole $vn_5w_7w_2n_4v$ appears, a contradiction. So, we have that
$$\mbox{$N_4$ is complete to $N_5$.}$$

Let $n_6\in N_6$. Note that $W_3$ is anticomplete to $W_6$ by Definition~\ref{def-hepta}$(2)$. Since $N_4$ is complete to $N_5$, and since  the vertices in $W_3\cup W_4\cup W_5\cup W_6\cup \{v\}$ cannot induce a full house, we know that $n_6$ is anticomplete to $N_5$. So, $n_6$ has a neighbor $w_5\in M_5$ by Definition~\ref{def-hepta}$(3)$, and $W_3$ is complete to $W_5$ by Lemma~\ref{common path}. Now, a 5-hole of form $v-N_5-W_3-w_5-n_6-v$ appears. This contradiction shows that $6\notin I$. Similarly, $3\notin I$.

Now, we may suppose $I=\{1, 2, 4, 5\}$ by symmetry. Let $W'=(W_1, W_2, W_3\cup \{v\}, W_4, W_5, W_6,$ $W_7)$. We will deduce a contradiction by showing  that  $G[W\cup \{v\}]$ is also a heptagram.

It is easy to see that $W'$ satisfies Definition~\ref{def-hepta}$(1\sim 3)$. From the symmetry, one sees also that:
\renewcommand{\baselinestretch}{1}
\begin{itemize}
	\item [$(i)$] Definition~\ref{def-hepta}$(4)$ $\Leftrightarrow$ $N_2$ is complete to $N_4$, and $N_4$ is anticomplete to $M_5$
	
	\item [$(ii)$] Definition~\ref{def-hepta}$(5)$ $\Leftrightarrow$ $M_2$ is anticomplete to $M_4$, and $M_4$ is complete to $N_5$.
	
	\item [$(iii)$] Definition~\ref{def-hepta}$(6)$ $\Leftrightarrow$ every vertex in $W_6$ is either anticomplete to $M_4$ or complete to $N_5$, and if $M_2\ne \emptyset$ then $N_4$ is complete to $N_5$.
\end{itemize}\renewcommand{\baselinestretch}{1.2}

To prove that $W'$ satisfies Definition~\ref{def-hepta}$(4, 5, 6)$, we just need to verify the right side of $(i),(ii),(iii)$.

Let $n_2\in N_2$, $n_4\in N_4$, and $w_5\in M_5$. To prove the right side of $(i)$, we may first suppose  that $n_2n_4\notin E(G)$. By Definition~\ref{def-hepta}$(3)$ and Lemma~\ref{common path}, there exist $w_6\in W_6$ and $w_7\in W_7$ such that $\{n_4w_6, n_2w_7, w_6w_7\}\subseteq E(G)$, which implies a 5-hole $vn_4w_6w_7n_2v$. This contradiction shows that $N_2$ is complete to $N_4$.
Now, we suppose that  $n_4w_5\in E(G)$. Let $n_1\in N_1$, we can find $w_7'\in N_{W_7}(n_1)\cap N(w_5)$, which implies a 5-hole  $vn_1w'_7w_5n_4v$. Therefore, the right side of $(i)$ holds.

Let $w_2\in M_2$, $w_4\in M_4$ and $n_5\in N_5$. To prove the right side of $(ii)$, we may first suppose $w_2w_4\in E(G)$. If $N(w_2)\cap N_1\ne\emptyset$ and $N(w_4)\cap N_5\ne \emptyset$, then there exists $n_1\in N_1$ and $n_5\in N_5$ such that  $vn_1w_2w_4n_5v$ will be a 5-hole.
So, suppose $N(w_2)\cap N_1=\emptyset$ and $n_1w_2\notin E(G)$.
By Definition~\ref{def-hepta}$(3, 4, 5)$, there exist $w_1\in W_1$ and $w_7\in W_7$ such that $\{w_1w_2,w_1w_7,w_2w_7\}\subseteq E(G)$ and $n_1w_7\notin E(G)$. Since $n_1w_2\notin E(G)$, we know that $n_4w_2\in E(G)$ by Lemma~\ref{common path}, which implies a 5-hole $vn_1w_7w_2n_4v$. This contradiction shows that $M_2$ is anticomplete to $M_4$.
Now, we suppose that $w_4n_5\notin E(G)$, then there exist $w_6\in W_6$ such that $w_4w_6\in E(G)$ by Definition~\ref{def-hepta}$(3)$ and $w_6n_5\in E(G)$ by Definition~\ref{def-hepta}$(5)$. Since $w_4n_5\notin E(G)$, we know that $W_4$ is not complete to $W_5$, and then by Lemma~\ref{common path}, we have that $W_2$ is complete to $W_4$ (in $W$), which implies that $n_2w_4\in E(G)$. So, there is a 5-hole $vn_2w_4w_6n_5v$. Therefore, the right side of $(ii)$ holds.

Let $w_6\in W_6$, $w_4\in M_4$, $n_5\in N_5$ and $n_4\in N_4$. To prove the right side of $(iii)$, we first suppose $w_6w_4\in E(G)$ and $w_6n_5\notin E(G)$. Then $n_5w_4\in E(G)$ by Definition~\ref{def-hepta}$(5)$.
Since $w_6n_5\notin E(G)$, we know that $w_6n_1\in E(G)$ by Lemma~\ref{common path}, which implies a 5-hole $vn_1w_6w_4n_5v$. This contradiction shows that every vertex in $W_6$ is either anticomplete to $M_4$ or complete to $N_5$.
Now, we may suppose $M_2\ne \emptyset$ and $n_4n_5\notin E(G)$. Then $w_2n_4\in E(G)$ by Lemma~\ref{common path}. Again by Lemma~\ref{common path}, there exists a vertex $w_7\in N_{W_7}(w_2)\cap N(n_5)$, which implies a 5-hole $vn_4w_2w_7n_5v$. Therefore, the right side of $(iii)$ holds, which implies that $(\ref{eqa-Y-vertex-4})$ holds.

\medskip

We will show that if $|I|=3$ then $v$ is a $Y$-vertex, and if $|I|=2$ then the second conclusion holds.

\begin{equation}\label{eqa-Y-vertex-5}
\mbox{If $|I|=3$, then $v$ is a $Y$-vertex.}
\end{equation}

We may first suppose that $I=\{1, 2, 3\}$. Let $n_i\in N_i$ for $i\in I$ and $w_4\in W_4$. We suppose by symmetry that $n_1n_2\notin E(G)$, as $G[\{n_1,n_2,n_3,w_4,v\}]$ is not a full house. Let $w_6\in W_6$, now a 5-hole $vn_1w_6w_4n_2v$ appears, a contradiction. So, $I\ne \{1, 2, 3\}$.

We may assume that $1, 4\in I$.
If $N_5=\emptyset$ and $N_7=\emptyset$, then there will be a hole of form $v-N_4-W_5-W_7-N_1-v$ by Lemma~\ref{common path}. So, $N_5\ne \emptyset$ or $N_7\ne \emptyset$, and then we may suppose that $I=\{1, 4, 5\}$ by symmetry.
By the same argument, $N_4$ is anticomplete to $M_5$ and $N_5$ is anticomplete to $M_4$.
If $N_4$ is not complete to $N_5$, then there is a 5-hole of the form $v-N_5-W_7-W_2-N_4-v$ by Lemma~\ref{common path}. So, $N_4$ is complete to $N_5$.
Suppose that $N_1$ is not complete to $W_2$.
Let $w_2\in W_2$ and $w_7\in W_7$ such that $n_1w_2\notin E(G)$ and $w_7w_2\in E(G)$ by Definition~\ref{def-hepta}$(3)$. Then $n_1w_7\in E(G)$ by Definition~\ref{def-hepta}$(5)$. Since $n_1w_2\notin E(G)$, we know that $w_2n_4\in E(G)$, which implies a 5-hole $vn_1w_7w_2n_4v$, a contradiction.
So, $N_1$ is complete to $W_2$, and $N_1$ is complete to $W_3$ by Definition~\ref{def-hepta}$(3, 4)$.
Similarly, $N_1$ is complete to $W_6\cup W_7$.
But then $v$ is a $Y$-vertex of type 1.
This proves (\ref{eqa-Y-vertex-5}).

\medskip

Now, we suppose $|I|=2$, and show that
\begin{equation}\label{eqa-Y-vertex-6}
\mbox{the second outcome of the lemma holds.}
\end{equation}

Let $I=\{1, t\}$, where $t\in \{2, 3, 4\}$. If $t=4$, there will be a 5-hole of form $v-N_4-W_5-W_7-N_1-v$, a contradiction. Suppose there exist $n_1\in N_1$ and $n_t\in N_t$ ($t\in \{2,3\}$) such that $n_1n_t\notin E(G)$. Let $w_6\in W_6$, then there exists $w_4\in N_{W_4}(n_t)\cap N(w_6)$ by Lemma~\ref{common path}, which implies a 5-hole $vn_1w_6w_4n_tv$. Thus, $N_1$ is complete to $N_t$ and (\ref{eqa-Y-vertex-6}) holds.

\medskip

From (\ref{eqa-Y-vertex-1})-(\ref{eqa-Y-vertex-6}), we prove Lemma~\ref{Y-ver}. \qed

\medskip

We will  prove further that no two $Y$-vertices of the same type may be adjacent.

\begin{lemma}\label{nonadj y}
For each $i\in [7]$, no two $Y$-vertices of type $i$ are adjacent.
\end{lemma}
\pf Suppose to its contrary, let  $a, b$ be two adjacent $Y$-vertices of type $5$.
For $j\in\{1, 2, 5\}$, let $A_j=N_{W_j}(a)$ and $B_j=N_{W_j}(b)$.

If $A_5\cap B_5=\emptyset$, choose $a_5\in A_5$ and $b_5\in B_5$,  since $W_5$ is a stable set we know that $\{a_5b_5, ab_5, a_5b\}\cap E(G)=\emptyset$, and since both $a$ and $b$ are $Y$-vertices of type 5 we see that $W_3\cup W_4\cup W_6\cup W_7$ has a common neighbor of $a_5$ and $b_5$, say $c$, then a 5-hole $ca_5abb_5c$ appears. Therefore, $A_5\cap B_5\ne\emptyset$. Let $n_5\in A_5\cap B_5$.

Now, we claim that $A_1\cap B_1=A_2\cap B_2=\emptyset$.
If it is not the case, we suppose by symmetry that $A_1\cap B_1\ne\emptyset$. Following Definition~\ref{def-y-vertex}, we know that $A_1$ is complete to $A_2$ and $B_1$ is complete to $B_2$, and so $G[A_1\cup A_2]$ and $G[B_1\cup B_2]$ are both components of $G[W_1\cup W_2]$.
Since $A_1\cap B_1\ne\emptyset$, we see that $A_1\cup B_1$ is complete to $A_2\cup B_2$, which implies that $A_2\cap B_2\ne\emptyset$. Let $n_1\in A_1\cap B_1$ and $n_2\in A_2\cap B_2$. Then $G[\{n_5, n_1, n_2, a, b\}]$ is a full house. This contradiction proves our claim.

Since $A_1\cap B_1=A_2\cap B_2=\emptyset$, we have that $A_1$ is anticomplete to $B_2$, which implies that $W_1$ is not complete to $W_2$.
Since $W_1$ is complete to $W_6$ by Lemma~\ref{common path}, by choosing $a_1\in A_1$, $b_1\in B_1$ and $w_6\in W_6$, we have a 5-hole $w_6a_1abb_1w_6$. This contradiction proves Lemma~\ref{nonadj y}.  \qed


\subsection{$V$-vertex}

In this subsection, we study the properties of  $V$-vertices defined  below (see also \cite{CRST10}).

\begin{define}\label{def-tail-t}
Let $t\in [7]$. A \textit{tail}, or \textit{tail of type t}, is an induced path $v_1v_2\ldots v_k$ with the following properties:

\begin{itemize}
	\item [$1.$] \noindent $k$ is odd, and $\{v_1, \ldots, v_k\}\subseteq V(G)\setminus W$,
	\item [$2.$] \noindent $v_1$ has neighbors in both $W_{t-3}$ and $W_{t+3}$, and $W_{t-3}\cup W_{t+3}$ is anticomplete to $\{v_2, \ldots, v_k\}$,
	\item [$3.$] \noindent $W_{t-1}\cup W_{t+1}$ and one of $W_{t-2}$ and $W_{t+2}$ are anticomplete to $\{v_1, \ldots, v_k\}$,
	\item [$4.$] \noindent $v_k$ has a neighbor in $W_t$, and $W_t$ is anticomplete to $\{v_1, \ldots, v_{k-1}\}$,
	\item [$5.$] \noindent  $N_{W_{t-3}}(v_1)$ is complete to $N_{W_{t+3}}(v_1)$ and anticomplete to $W_{t+3}\setminus N_{W_{t+3}}(v_1)$, and $N_{W_{t+3}}(v_1)$ is anticomplete to $W_{t-3}\setminus N_{W_{t-3}}(v_1)$,
	\item [$6.$] \noindent $N_{W_t}(v_k)$ is complete to $W_{t-2}\cup W_{t-1}\cup W_{t+1}\cup W_{t+2}$.
\end{itemize}
\end{define}

Following from Lemma~\ref{Y-ver} and the definition above, we see that
\begin{equation}\label{0 length V-ver}
\mbox{every $Y$-vertex forms a tail of length zero.}
\end{equation}


We call a tail $v_1v_2\ldots v_k$ as a \textit{tail for $v_1$}. For $t\in [7]$, a vertex $v\in V(G)\setminus W$ is called a \textit{hat of type $t$} if it has neighbors in both $W_{t-3}$ and $W_{t+3}$, and is anticomplete to $W\setminus (W_{t-3}\cup W_{t+3})$. If $k\ge 2$ and $v_1v_2\ldots v_k$ is a tail of type $t$, then $v_1$ is a hat of type $t$. We say that a vertex $v\in V(G)\setminus W$ is a \textit{$V$-vertex of type $t$} if there is a tail of type $t$ for $v$. Thus, every $V$-vertex of type $t$ is either a $Y$-vertex of type $t$ or a hat of type $t$.

Below Lemmas~\ref{W_-1 0 +1}, \ref{adj-Y}, \ref{V-ver not com},  \ref{i,i+1 adj},  \ref{V-ver not com and no type V}, and \ref{remaining} and their proofs are respectively the same as 7.1, 7.2, 7.4, 7.5, 7.6, and 8.1 of \cite{CRST10}. We leave their proofs in the appendix for referees.
\begin{lemma}\label{W_-1 0 +1}
Let $X\subseteq V(G)\setminus W$, such that $G[X]$ is connected and contains no tail of $G$. Then there exists $i\in [7]$ such that $N(X)\cap W\subseteq W_{i-1}\cup W_i\cup W_{i+1}$.
\end{lemma}

\begin{lemma}\label{adj-Y}
Let $U$ be the set of all vertices in $V(G)\setminus W$ that are not $V$-vertices. For $t\in [7]$, if an induced  path $x_1x_2\cdots x_k$ satisfies the following:
	\begin{itemize}
		\item \noindent $x_1$ is either a hat or $Y$-vertex of type $t$,
		\item \noindent $\{x_2, \ldots, x_{k-1}\}\subseteq U$, and $x_k\in N(W_{t+1}\cup W_{t-1})\setminus W$,
	\end{itemize}
    then  $x_k$ is a $Y$-vertex of type $t+1$ or $t-1$. 
\end{lemma}

\begin{lemma}\label{V-ver not com}
Let $i\in[7]$, and let $x$ be a $V$-vertex of type $i$. If $x$ is not complete to $W_{i-3}\cup W_{i+3}$, then $W_{i-2}\cup W_{i+2}$ is complete to $W_{i-3}\cup W_{i+3}$.
\end{lemma}

\begin{lemma}\label{i,i+1 adj}
Let $i\in [7]$. If $u$ is a $V$-vertex of type $i$, and $v$ is a $V$-vertex of type $i+1$, then $uv\in E(G)$ and both $u$ and $v$ are complete to $W_{i-3}$.
\end{lemma}

\begin{lemma}\label{V-ver not com and no type V}
Let $i\in [7]$. If $x$ is a $V$-vertex of type $i$, and $x$ is not complete to $W_{i-3}\cup W_{i+3}$, then there is no $V$-vertex of type $j$ for $j\in \{i-3, i-1, i+1, i+3\}$.
\end{lemma}

\begin{lemma}\label{remaining}
Every vertex in $V(G)\setminus W$ is a $V$-vertex.
\end{lemma}

Now, we are ready to prove Theorem~\ref{main'}.

\pf Let $G$ be an (odd hole, full house)-free graph, contains no harmonious cutset, and contains an antihole on seven vertices. By Lemma~\ref{T11}, $G$ is $T_{11}$-free. So, we choose a maximal heptagram, $W=(W_1, \dots, W_7)$ in $G$.
By Lemma~\ref{remaining}, every vertex of $G$ either belongs to $W$ or is a $V$-vertex.
For $i\in [7]$, let $Y_i$ be the set of all $Y$-vertices of type $i$.
By Lemma~\ref{nonadj y}, $Y_i$ is a stable set.
We need to check the requirements of Definition~\ref{def-hepta-type}.

The first is clear. The second and the third follows from Lemma~\ref{common path}.
The fourth to the seventh follows from Definitions~\ref{def-hepta} and \ref{def-y-vertex}.
Lemma~\ref{adj-Y} and Lemma~\ref{i,i+1 adj} imply the eighth, and the ninth follows from Lemma~\ref{V-ver not com} and Lemma~\ref{V-ver not com and no type V}.

For the last one, choose $y_{i-1}\in Y_{i-1}$, $y_i\in Y_i$, $y_{i+1}\in Y_{i+1}$, then $\{y_iy_{i-1}, y_iy_{i+1}\}\subseteq E(G)$ and $y_{i-1}y_{i+1}\notin E(G)$.
But now $y_i-y_{i+1}-W_{i+1}-W_{i-1}-y_{i-1}-y_i$ is an odd hole, a contradiction.
This complete the proof of Theorem~\ref{main'}. \qed

\pagebreak

\newpage

\appendix
\noindent{Appendix: The proofs of Lemmas~\ref{W_-1 0 +1}, \ref{adj-Y}, \ref{V-ver not com},  \ref{i,i+1 adj},  \ref{V-ver not com and no type V}, and \ref{remaining}.}

\begin{lem37}
Let $X\subseteq V(G)\setminus W$, such that $G[X]$ is connected and contains no tail of $G$. Then there exists $i\in [7]$ such that $N(X)\cap W\subseteq W_{i-1}\cup W_i\cup W_{i+1}$.
\end{lem37}
\pf Suppose to its contrary, and let $X$ be a minimal counterexample. For $i\in[7]$, let $N_i=N(X)\cap W_i$. Then there exists $i\in [7]$ such that none of $N_i$ and $N_{i+3}$ is anticomplete to $X$. We may therefore assume that $N_1\ne\emptyset$ and $N_4\ne\emptyset$.
Choose $n_4v_1v_2\ldots v_kn_1$ to be a minimal path from $W_4$ to $W_1$ with interior in $X$, where $n_1\in W_1$ and $n_4\in W_4$, and let $P=v_1v_2\ldots v_k$.
By the minimality of $X$ and Lemma~\ref{Y-ver}, it follows that $X=\{v_1, \ldots, v_k\}$ with $k>1$, $W_1$ is anticomplete to $X\setminus \{v_k\}$ and $W_4$ is anticomplete to $X\setminus \{v_1\}$.

We claim first that $k$ is odd. If it is not the case, suppose that $k$ is even. By Lemma~\ref{common path}, we know that $n_1$ and $n_4$ share at least one common neighbor $w_j\in W_j$ for each $j\in \{2, 3, 6\}$, it follows that no vertex of $\{w_2, w_3, w_6\}$  is anticomplete to $\{v_1, v_2, \ldots, v_k\}$ since $n_4-P-n_1w_jn_4$ cannot be an odd hole. By Lemma~\ref{Y-ver},  each of $v_1$ and $v_k$ is nonadjacent to one of $w_2$ and $w_3$, and consequently one of $w_2$ and $w_3$ is joined to $w_6$ by a path with interior a proper subpath of $v_1v_2\ldots v_k$, contrary to the minimality of $X$. So, $k$ is odd.

Since there is no odd hole of form $n_4v_1 \ldots v_kn_1-W_7-W_5-n_4$, we know that $W_5\cup W_7$ are not anticomplete to $X$, and we may assume by symmetry  that $W_5$ has a neighbor in  $X$. By the minimality of $X$, $X\setminus \{v_1\}$ is anticomplete to $W_5$. So, $N_{W_5}(v_1)\neq \emptyset$. Note that $G[X]$ contains no tail and thus neither $Y$-vertex by (\ref{0 length V-ver}). We see that $v_1$ must be a hat of type 1 by Lemma~\ref{Y-ver}. Next, we will deduce a contradiction by showing that  $P$ is a tail.

By the minimality of $X$, $W_2\cup W_7$ is anticomplete to $X\setminus \{v_k\}$.
Suppose that $v_k$ has a neighbor, say $n_2$, in $W_2$. Then $v_k$ is a hat of type 5 by Lemma~\ref{Y-ver}, and so $v_k$ is anticonplete to $W_7$, which implies that $W_7$ is anticomplete to $X$. From the minimality of $X$, and since $v_k$ has a neighbor in $W_2$, we see that $W_6$ is anticomplete to $X$ too.
If $n_2n_4\in E(G)$, then $n_4-P-n_2n_4$ is an odd hole, and so $n_2n_4\notin E(G)$.
By Lemma~\ref{common path}, there exist $w_6\in W_6$ and $w_7\in W_7$ such that $n_4-P-n_2w_7w_6n_4$ is an odd hole. This contradiction shows that $v_k$ is anticonplete to $W_2$. Therefore,
$$\mbox{$X$ is anticomplete to $W_2$, and to $W_7$ similarly.}$$

Since $v_1$ is a hat of type 1, we know that $v_1$ is anticomplete to $W_3\cup W_6$ by Lemma~\ref{Y-ver}. By the minimality of $X$, either  $W_3$ or $W_6$ is anticomplete to $X\setminus \{v_1\}$, and so is anticomplete to $X$.

Now, we have checked that $P$ satisfies the first four requirements in the definition of tails.

To verify the fifth requirement, let $S_i=N_{W_i}(v_1)$ and $T_i=W_i\setminus S_i$ for $i=4,5$. Since $v_1$ is a hat of type 1, we know that $S_4$ is complete to $S_5$ by Lemma~\ref{Y-ver}. If $w_4w_5\in E(G)$ for $w_4\in S_4$ and $w_5\in T_5$, then by Lemma~\ref{common path}, one can find a vertex $w_7'\in W_7$ such that $w_4-P-n_1w_7'w_5w_4$ is an odd hole. A similar contradiction occurs if $S_5$ is not anticomplete to $T_4$.  Therefore, $S_4$ is anticomplete to $T_5$, and $S_5$ is anticomplete to $T_4$ too.

To verify the last requirement, we choose $n_1\in N_{W_1}(v_k)$. If $n_1w_2\not\in E(G)$ for some  $w_2\in W_2$, then there exist $w_7''\in W_7$ such that $w_2w_7''\in E(G)$ by Definition~\ref{def-hepta}$(3)$, and consequently $n_1w_7''\in E(G)$ by Definition~\ref{def-hepta}$(5)$.
By Lemma~\ref{common path}, we see that $W_2$ is complete to $W_4$, which implies an odd hole $n_4-P-n_1w_7''w_2n_4$. So, $N_{W_1}(v_k)$ is complete to $W_2$.
Since $W_2$ is complete to $W_3$ by Lemma~\ref{common path}, we have $N_{W_1}(v_k)$ is complete to $W_3$ by Definition~\ref{def-hepta}$(4)$, and similarly $N_{W_1}(v_k)$ is complete to $W_6\cup W_7$.

Now, $P$ is a tail in $G[X]$. This contradiction proves Lemma~\ref{W_-1 0 +1}. \qed

\begin{lem38}
Let $U$ be the set of all vertices in $V(G)\setminus W$ that are not $V$-vertices. For $t\in [7]$, if an induced  path $x_1x_2\cdots x_k$ satisfies the following:
	\begin{itemize}
		\item \noindent $x_1$ is either a hat or $Y$-vertex of type $t$,
		\item \noindent $\{x_2, \ldots, x_{k-1}\}\subseteq U$, and $x_k\in N(W_{t+1}\cup W_{t-1})\setminus W$,
	\end{itemize}
    then  $x_k$ is a $Y$-vertex of type $t+1$ or $t-1$. 
\end{lem38}
\pf Suppose that a path $P=x_1x_2\ldots x_k$ is a counterexample with $k$ minimum, and let $X=\{x_1, \ldots, x_k\}$. Without loss of generality, we suppose that $t=1$, and suppose that  $x_k$ has a neighbor in $W_2$. From the minimality of $k$,
\begin{equation}\label{W2-W7-anticomplete}
\mbox{$W_2\cup W_7$ is anticomplete to $X\setminus\{x_k\}$.}
\end{equation}
Let $w_2\in N_{W_2}(x_k)$. We claim that
$$\mbox{there exists $w_4\in N_{W_4}(x_1)$ such that $w_2w_4\in E(G).$}$$
Since $x_1$ is a $V$-vertex, if $W_4$ is complete to $W_5$, then $x_1$ is complete to $W_4$, which implies the existence of a required $w_4$ by Definition~\ref{def-hepta}$(3)$. If $W_4$ is not complete to $W_5$, then $W_4$ is complete to $W_2$ by Lemma~\ref{common path}. Therefore, the claim holds. 

\medskip


Next, we show that
\begin{equation}\label{x_1 and x_k tail}
\mbox{$G[X]$ contains a tail for $x_k$ and a tail for $x_1$.}
\end{equation}

Suppose that $G[X]$ contains no tail for $x_k$, i.e., $x_k$ is not a $V$-vertex. Since $x_1$ is either a hat or $Y$-vertex of type $t$, $G[X]$ contains a tail for $x_1$ and  it contains no tail for $x_k$. Since $N_{W_2}(x_k)\neq \emptyset$, and since $\{x_2, \ldots, x_{k-1}\}\subseteq U$, we deduce that $W_5\cup W_6$ is anticomplete to $X\setminus\{x_1\}$ by Lemma~\ref{W_-1 0 +1}, and then $W_6$ is anticomplete to $X$.
As $G[X]$ contains a tail for $x_1$, there exists $j\le k$, such that $x_1\dots x_j$ is a tail for $x_1$. By the definition of tail, we know that $W_2$ is anticomplete to $\{x_1,\dots, x_j\}$, which implies that $j<k$.

Suppose that $k$ is even. Since $W_5\cup W_6$ is anticomplete to $X\setminus\{x_1\}$ and there is no odd hole of the form $x_1Px_kw_2-W_7-W_5-x_1$, we see that $x_k$ has a neighbor $w_7\in W_7$ by the minimality of $X$, and consequently $W_4$ is anticomplete to $X\setminus \{x_1\}$ by Lemma~\ref{W_-1 0 +1}, which implies an odd hole of the form $x_1Px_kw_7-W_6-W_4-x_1$ by Lemma~\ref{common path}. This contradiction shows that $k$ must be odd.
Since $x_1Px_kw_2w_4x_1$ is not an odd hole, we can deduce that $w_4$ has a neighbor in $X\setminus \{x_1\}$ by (\ref{W2-W7-anticomplete}).
By applying Lemma~\ref{W_-1 0 +1} to $X\setminus\{x_1\}$, we have that $W_1$ are anticomplete to $X\setminus \{x_1\}$ and therefore $j=1$. So, $x_1$ is a $Y$-vertex.

Let $w_1\in N_{W_1}(x_1)$. Then $w_1$ is complete to $W_2$ by Definition~\ref{def-y-vertex}. Particularly,  $w_1w_2\in E(G)$, which implies an odd hole $x_1Px_kw_2w_1x_1$. This contradiction shows that $G[X]$ must contain a tail for $x_k$, and thus $x_k$ is either a hat or a $Y$-vertex of type $s$, where $s\in \{5, 6\}$. As $x_1$ has a neighbor in $W_{s-1}$, there is symmetry between $x_1$ and $x_k$, it follows that $G[X]$ also contains a tail for $x_1$. Therefore, (\ref{x_1 and x_k tail}) holds.

\medskip

Then we  show that
\begin{equation}\label{x_k not 5 or 6}
\mbox{$x_k$ is not a $V$-vertex of type 5 or 6,}
\end{equation}
and thus deduce a contradiction.

We first suppose that $x_k$ is a $V$-vertex of type 6. Then $x_k$ has a neighbor in $W_3$.
By the minimality of $k$ and by Lemma~\ref{W_-1 0 +1}, we know that $W_7$ is anticomplete to $X$, $W_2$ is anticomplete to $X\setminus \{x_k\}$, and $W_5$ is anticomplete to $X\setminus \{x_1\}$.
If $k$ is even then an odd hole of the form $x_1Px_kw_2-W_7-W_5-x_1$ appears. This shows that $k$ must be odd.
Since $w_4x_1Px_kw_2w_4$ is not an odd hole, it follows that $w_4$ has a neighbor in $X\setminus\{x_1, x_k\}$.
By applying Lemma~\ref{W_-1 0 +1} to $X\setminus\{x_1, x_k\}$, we see that $W_1$ is anticomplete to $X\setminus\{x_1, x_k\}$. But then, $W_1$ has a vertex, say $w_1$, by (\ref{x_1 and x_k tail}) such that $N_{X}(w_1)$ has a vertex which is in a tail for $x_1$. Since $x_k$ is a $V$-vertex of type 6, we see that $w_1x_k\notin E(G)$, and so $w_1$ is nonadjacent to $X\setminus\{x_1\}$. In particular, $w_1x_1\in E(G)$. Since $x_1$ is a $V$-vertex, we know that $w_1$ is complete to $W_2$, and so there exists $w'_2\in N_{W_2}(w_1)$ which implies an odd hole $w_1x_1Px_kw'_2w_1$, a contradiction.

So, we suppose that $x_k$ is a $V$-vertex of type 5. Then $x_k$ has a neighbor in $W_1$, and $W_4\cup W_6$ is anticomplete to $X\setminus\{x_1\}$ by the minimality of $k$. If $k$ is odd then $x_1Px_kw_2w_4x_1$ is an odd hole. Therefore, $k$ is even.
Let $w_5\in N_{W_5}(x_1)$, $w_1\in N_{W_1}(x_k)$, and $w_6\in N_{W_6}(w_5)\cap N_{W_6}(w_1)$ by Lemma~\ref{common path}. Since $x_1Px_kw_1w_6w_5x_1$ is not an odd hole, we see that $w_5$ is not anticomplete to $X\setminus \{x_1\}$. Similarly, $w_1$ is not anticomplete to $X\setminus\{x_k\}$.
By applying Lemma~\ref{W_-1 0 +1} to $X\setminus\{x_1, x_k\}$, either $w_1$ or $w_5$ has no neighbor in $X\setminus\{x_1,x_k\}$. By symmetry we may suppose that $w_1x_1\in E(G)$ and $w_1$ is anticomplete to $X\setminus\{x_1, x_k\}$. Then $x_1$ is a $Y$-vertex.
Since $x_1Px_kw_1x_1$ is not an odd hole, it follows that $k=2$, and so $w_5x_2\in E(G)$, which implies that $x_k$ is also a $Y$-vertex.

Since $x_1$ is a $Y$-vertex, we see that $N_{W_1}(x_1)$ is complete to $W_2$, and then $G[W_1\cup W_2]$ is connected by Definition~\ref{def-hepta}$(3)$.
Since $x_2$ is a $Y$-vertex of type 5, $N_{W_1}(x_2)$ is complete to $N_{W_2}(x_2)$.
Consequently $W_1$ is complete to $W_2$ and $x_2$ is complete to $W_1\cup W_2$.
Similarily, $x_1$ is complete to $W_4\cup W_5$ and $W_4$ is complete to $W_5$.
If $x_1$ is not complete to $W_1$, let $w_1$ be a non-neighbor of $x_1$ in $W_1$, then an odd hole of the form $x_1x_2w_1-W_3-w_4x_1$ appears. So, $x_1$ is complete to $W_1$, and similarly $x_2$ is complete to $W_5$.

Define $W_6'=W_6\cup \{x_1\}$ and $W_7'=W_7\cup \{x_2\}$. Let $W'=\{W_1,W_2,\dots,W_6',W_7'\}$, then one can easily check that $W'$ is heptagram a larger  than $W$. This contradiction shows that $x_k$ is not a $V$-vertex of type 5. Therefore, (\ref{x_k not 5 or 6}) holds.

\medskip

Since $N_{W_2}(x_k)\neq \emptyset$ and $x_k$ is a $V$-vertex by (\ref{x_1 and x_k tail}), we see that $x_k$ must be a $Y$-vertex of type 2 by (\ref{x_k not 5 or 6}) and Lemma~\ref{Y-ver}, i.e., Lemma~\ref{adj-Y} holds. \qed

\begin{lem39}
Let $i\in[7]$, and let $x$ be a $V$-vertex of type $i$. If $x$ is not complete to $W_{i-3}\cup W_{i+3}$, then $W_{i-2}\cup W_{i+2}$ is complete to $W_{i-3}\cup W_{i+3}$.
\end{lem39}
\pf Without loss of generality, we assume that $i=1$. For $j\in\{4, 5\}$, let $N_j=N_{W_j}(x)$ and $M_j=W_j\setminus N_j$. By Definition~\ref{def-tail-t}, we see that $M_4\cup M_5\ne \emptyset$, $N_4$ is complete to $N_5$, $N_4$ is anticomplete to $M_5$, and $M_4$ is anticomplete to $N_5$. Then $M_4$ is linked to $M_5$ by Definition~\ref{def-hepta}$(3)$, which also implies that $M_4\ne \emptyset$ and $M_5\ne \emptyset$.

We will show that $W_6$ is complete to $W_4\cup W_5$. Let $w_6\in W_6$.
If $w_6$ is anticomplete to $M_4$, then $w_6$ has a neighbor in $N_4$ by Definition~\ref{def-hepta}$(3)$, and $w_6$ is complete to $M_5$ by Definition~\ref{def-hepta}$(5)$. Since $M_4$ and $M_5$ are linked and by Definition~\ref{def-hepta}$(4)$, we see that $w_6$ is complete to $M_4$, a contradiction. Therefore, there exists an $m_4\in N_{M_4}(w_6)$. Since $m_4$ is anticomplete to $N_5$, we see that $w_6$ is complete to $N_5$ by Definition~\ref{def-hepta}$(5)$, and consequently $w_6$ is complete to $N_4$ by Definition~\ref{def-hepta}$(4)$.
Let $n_4\in N_4$. Since $n_4$ is anticomplete to $M_5$, we know that $w_6$ is complete to $M_5$ by Definition~\ref{def-hepta}$(5)$, and hence complete to $M_4$ by Definition~\ref{def-hepta}$(4)$.
This proves that $w_6$ is complete to $W_4\cup W_5$, and so is $W_6$. Similarly, we can show that $W_3$ is complete to $W_4\cup W_5$. This proves Lemma~\ref{V-ver not com}. \qed


\begin{lem310}
Let $i\in [7]$. If $u$ is a $V$-vertex of type $i$, and $v$ is a $V$-vertex of type $i+1$, then $uv\in E(G)$ and both $u$ and $v$ are complete to $W_{i-3}$.
\end{lem310}
\pf Without loss of generality, we suppose that $i=1$. Let $u$ and $v$ be $V$-vertices of type 1 and 2, and let their tails be $S$ and $T$, respectively, such that both $S$ and $T$ have minimum length among all feasible choices. It is certain that both $S$ and $T$ have odd number of vertices. Let $A_j=N_{W_j}(u)$ for $j\in \{4, 5\}$, and let $B_j=N_{W_j}(v)$ for $j\in \{5, 6\}$.
If $u$ is not complete to $W_5$, then $W_3\cup W_6$ is complete to $W_4\cup W_5$ by Lemma~\ref{V-ver not com}, and consequently $v$ must be complete to $W_5$ by Definition~\ref{def-tail-t}(2, 5). Therefore,
\begin{equation}\label{one-of-u-v}
\mbox{one of $u$ and $v$ is complete to $W_5$, and $N_{W_5}(u)\cap N_{W_5}(v)\neq\emptyset$.}
\end{equation}

We first suppose that $uv\notin E(G)$. By (\ref{one-of-u-v}), we may choose a vertex $w_5\in N_{W_5}(u)\cap N_{W_5}(v)$.
If $S$ and $T$ are disjoint and $E(S, T)=\emptyset$, since both $S$ and $T$ have odd numbers of vertices, $G$ has an odd hole of the form $u-S-W_1-W_2-T-vw_5u$, a contradiction (note that $w_5$ has no neighbor in $S\cup T$ except $u$ and $v$).
Thus, $G[S\cup T]$ is connected.

Since $S$ is a tail for $u$ and $T$ is a tail for $v$, we see that $u$ is anticomplete to $V(T)\setminus \{v\}$ by Lemma~\ref{adj-Y}, and consequently $u$ is anticomplete to $V(T)$. Similarly, $v$ is anticomplete to $V(S)$.

Let $X=V(S)\cup V(T)\setminus \{u, v\}$. Since $G[S\cup T]$ is connected, it follows that $G[X]$ is connected and both $S$ and $T$ are of lengths greater than 0, and so $N(X)\cap W_1\ne \emptyset$ and $N(X)\cap W_2\ne \emptyset$.
By the minimality of $S$ and $T$, $X$ contains no $V$-vertex, we see that $W_4\cup W_6$ is anticomplete to $X$ by Lemma~\ref{W_-1 0 +1}.
Choose $a_4\in A_4$ and $b_6\in B_6$ such that $w_5a_4\in E(G)$ and $w_5b_6\in E(G)$. Then $a_4b_6\in E(G)$ by Definition~\ref{def-hepta}$(4)$, which implies that one of $G[X\cup \{w_5\}]$ and $G[X\cup \{a_4, b_6\}]$ contains an odd hole. This contradiction shows that
\begin{equation}\label{uv-in-EG}
uv\in E(G).
\end{equation}

Now we turn to prove that both $u$ and $v$ are complete to $W_5$. If not the case,  we suppose, by (\ref{one-of-u-v}) and by symmetry, that $u$ is complete to $W_5$ but $v$ is not, and choose $a_5\in W_5\setminus B_5$. Since $u$ is complete to $W_5$, we see that $a_5\in A_5=W_5$.
For each vertex $b_6\in B_6$, since $v$ is a $V$-vertex, then $a_5b_6\notin E(G)$ following Definition~\ref{def-tail-t}(5).
By Definition~\ref{def-hepta}$(3, 5)$, there exists $w_7\in N_{W_7}(a_5)$ such that $w_7b_6\in E(G)$, which implies an odd hole $uvb_6w_7a_5u$. This contradiction shows that both $u$ and $v$ are complete to $W_5$ and hence ends the proof of Lemma~\ref{i,i+1 adj}. \qed

\begin{lem311}
Let $i\in [7]$. If $x$ is a $V$-vertex of type $i$, and $x$ is not complete to $W_{i-3}\cup W_{i+3}$, then there is no $V$-vertex of type $j$ for $j\in \{i-3, i-1, i+1, i+3\}$.
\end{lem311}
\pf We may suppose that $i=1$. Let $x$ be a $V$-vertex of type 1, which is not complete to $W_4\cup W_5$. Since $W_4$ and $W_5$ are linked, we see that for each $m\in W_4\setminus N(x)$, there exists an $n\in N_{W_5}(m)$. By Definition\ref{def-tail-t}(5),  if $x$ is complete to $W_5$ then it is complete to $W_4$,  and vice versa. Therefore, $x$ is complete to neither $W_4$ nor $W_5$. By Lemma~\ref{i,i+1 adj}, there is no $V$-vertex of type 2 or of type 7.
Since $x$ is complete to neither $W_4$ nor to $W_5$, again by Definition~\ref{def-tail-t}(5), we see that no vertex in $W_4$ is complete to $W_5$, and consequently there is no $V$-vertex of type 4 by Definition~\ref{def-tail-t}. Similarly there is no $V$-vertex of type 5.
Lemma~\ref{V-ver not com and no type V} follows. \qed


Recall that $G$ has no  a harmonious cutset by our assumption. We will show that all  vertices in $V(G)\setminus W$ are $V$-vertices.

\begin{lem312}
Every vertex in $V(G)\setminus W$ is a $V$-vertex.
\end{lem312}
\pf Let $U$ be the set of all vertices of $V(G)\setminus W$ that are not $V$-vertices, and suppose that $U\ne\emptyset$. Choose  $X\subseteq U$ to be a maximum subset such that $G[X]$ is connected, and let $N(X)$ be the neighborhood of $X$ in $V(G)\setminus U$.  It is obvious that $X\ne\emptyset$ as $U\ne\emptyset$.

For $i\in [7]$, let $V_i$ be the set of all $V$-vertices of type $i$, and let $N_i=N(X)\cap W_i$ and $P_i=N(X)\cap V_i$. Note here the symbol $N_i$ represents a subset different from those in Definitions~\ref{def-hepta-type} and \ref{def-y-vertex}.

Let $I=\{i\in [7]\;:\; N_i\ne \emptyset\}$ and $J=\{i\in [7]\;:\;P_i\ne \emptyset\}$.
By Lemmas~\ref{W_-1 0 +1} and \ref{adj-Y}, there exist $s$ and $t$ such that
\begin{equation}\label{IJ}
	I\subseteq \{s-1, s, s+1\}, \mbox{ and }  J\subseteq \{t, t+1\}.
\end{equation}

\medskip

We first claim that for $i\in [7]$,
\begin{equation}\label{a b even path}
\mbox{if $a, b\in N_i$, then there is an induced even $ab$-path with interior in $X$.}
\end{equation}

Without loss of generality, we suppose that $i=1$. Then, $4\not\in I$ and $5\notin I$, and either $3\notin I$ or $6\notin I$. From the symmetry, we may assume that $6\notin I$.
Let $Q$ be an induced $ab$-path with interior in $X$.
If there exists a $w_6\in W_6\cap N(a)\cap N(b)$, (\ref{a b even path}) holds since $w_6aQbw_6$ is not an odd hole. So, we suppose that $W_6\cap N(a)\cap N(b)=\emptyset$. Then $W_6$ is complete to $W_5$ by Lemma~\ref{common path}. Let $a'\in N_{W_6}(a)\setminus N(b)$, $b'\in N_{W_6}(b)\setminus N(a)$, and $w_5\in W_5$. Then $w_5b'bQaa'w_5$ is a hole that must be of even length. Thus, (\ref{a b even path}) holds.

\medskip

Next, we claim that
\begin{equation}\label{Ni com to Ni+1}
\mbox{for $i\in [7]$, $N_i$ is complete to $N_{i+1}$.}
\end{equation}

Without loss of generality, suppose $n_1\in N_1$ and $n_2\in N_2$ such that $n_1n_2\notin E(G)$.
Let $Q$ be an induced $n_1n_2$-path with interior in $X$.
By (\ref{IJ}), $4\not\in I$ and $6\notin I$, and $3\notin I$ or $7\notin I$. We suppose $3\notin I$ and let $w_3\in N_{W_3}(n_1)$. Then $n_2w_3\in E(G)$ by Definition~\ref{def-hepta}$(5)$.
As $w_3n_1Qn_2w_3$ cannot be an odd hole, we see that $Q$ is of even length, which implies an odd hole of the form $n_1Qn_2-W_4-W_6-n_1$. Therefore, (\ref{Ni com to Ni+1}) holds.

\medskip

Now, we present two properties on $P_i$. For $i\in [7]$,
\begin{equation}\label{P_i same neigh}
	\mbox{two members of $P_i$ have the same neighbors in $W_{i-3}\cup W_{i+3}$,}
\end{equation}
and
\begin{equation}\label{P_i com to Ni-3 and Ni+3}
	\mbox{$P_i$ is complete to $N_{i-3}\cup N_{i+3}$.}
\end{equation}

Suppose that $i=1$ and $P_1\ne \emptyset$.
For $j\in\{4, 5\}$, let $R_j=N(X\cup P_1)\cap W_j$.
Suppose that $R_4$ is not complete to $R_5$. Let $r_4\in R_4$ and $r_5\in R_5$ with $r_4r_5\notin E(G)$, and let $Q$ be an induced $r_4r_5$-path with interior in $X\cup P_1$.
If $X\cup P_1$ is not anticomplete to $W_2\cup W_7$, then there exists a $Y$-vertex of type 2 or 7 by Lemma~\ref{adj-Y}, and consequently each vertex of $P_1$ is complete to $W_4$ or $W_5$ by Lemma~\ref{i,i+1 adj} (note that $P_1\ne\emptyset$), which implies that $W_4$ must be complete to $W_5$ by Definition~\ref{def-tail-t}. So, $X\cup P_1$ is anticomplete to $W_2\cup W_7$, and hence $X\cup P_1$ is anticomplete to at least one of $W_3$ and $W_6$ by lemma~\ref{W_-1 0 +1}. We suppose $X\cup P_1$ is anticomplete to $W_3$ by symmetry.
But now an odd hole occurs either of the form $r_4-W_3-r_5-Q-r_4$ or of the form $r_4-W_2-W_7-r_5-Q-r_4$.
This contradiction shows that $R_4$ is complete to $R_5$. It follows directly, by Definition~\ref{def-tail-t}, that (\ref{P_i same neigh}) holds.

Since each $p_1\in P_1$ is a $V$-vertex, by Definition~\ref{def-tail-t}, we see that $N_{W_4}(p_1)$ is complete to $N_{W_5}(p_1)$.
As $R_4$ is complete to $R_5$, it follows that each $p_1\in P_1$ is complete to $R_4\cup R_5$. This proves (\ref{P_i com to Ni-3 and Ni+3}).


\medskip

To complete the proof of Theorem~\ref{main'}, we will deduce a contradiction by constructing a harmonious cutset of $G$.  For doing this, we still need some properties of $I$ and $J$.

\begin{equation}\label{J ne empty}
J\ne\emptyset.
\end{equation}

Suppose $J=\emptyset$, and suppose that $I\subseteq \{1, 2, 3\}$ by symmetry.
By (\ref{Ni com to Ni+1}), $N_2$ is complete to $N_1$ and $N_3$.
It follows that $N_1$ is complete to $N_3$ by Definition~\ref{def-hepta}$(4)$ whenever $N_2\ne \emptyset$. By applying (\ref{a b even path}) and Lemma~\ref{harmonious set} to the cutset $N_1\cup N_2\cup N_3$, we deduce that $G$ admits a harmonious cutset.
This contradiction shows that $N_2=\emptyset$.
If $N_1$ is complete to $N_3$, by the same argument before, one can find a harmonious cutset in $G$. So, suppose $n_1\in N_1$ and $n_3\in N_3$ such that $n_1n_3\notin E(G)$.
Let $Q$ be an induced $n_1n_3$-path with interior in $X$. We just need to prove that $Q$ is of odd length. If $Q$ is not of odd length, then an odd hole of the form $n_1-Q-n_3-W_4-W_6-n_1$ appears by Lemma~\ref{common path}.
Therefore, (\ref{J ne empty}) holds.

\begin{equation}\label{I cap J = empty}
I\cap J=\emptyset.
\end{equation}

Suppose to its contrary and let $1\in I\cap J$ by symmetry.
By (\ref{IJ}), $\{4, 5\}\cap I=\emptyset$ and $\{3, 4, 5, 6\}\cap J=\emptyset$.
Since $X\subseteq U$ and $1\in J$, it follows from Lemma~\ref{adj-Y} that $\{2, 7\}\cap I=\emptyset=\{2, 7\}\cap J$ (as otherwise $\{4, 5\}\subseteq I$).
Consequently, $I\subseteq \{1, 3, 6\}$ and $J=\{1\}$. Again by (\ref{IJ}), we may assume that $I\subseteq \{1, 6\}$.
We will show that $P_1\cup N_6\cup N_1$ is a harmonious cutset with ($P_1\cup N_6$, $N_1$) being a partition of $P_1\cup N_6\cup N_1$ as required.
We need to check the following:

\begin{itemize}
	\item \noindent If $\{a, b\}\subseteq P_1\cup N_6$, then there is an induced even $ab$-path with interior disjoint from $P_1\cup N_6\cup N_1$,
	\item \noindent If $\{a, b\}\subseteq N_1$, then there is an induced even $ab$-path with interior disjoint from $P_1\cup N_6\cup N_1$, and
	\item \noindent If $a\in P_1\cup N_6$ and $b\in N_1$, then there is an induced odd $ab$-path with interior disjoint from $P_1\cup N_6\cup N_1$.
\end{itemize}

For the first , it follows directly from (\ref{a b even path}) if $\{a, b\}\subseteq N_6$, so we assume that $a\in P_1$, but then $N_{W_5}(a)\cap N_{W_5}(b)\neq\emptyset$ by Lemma~\ref{V-ver not com}, and the first statement follows since $W_5\cap (P_1\cup N_6\cup N_1)=\emptyset$.
The second statement follows directly from (\ref{a b even path}).
For the third, let $a\in P_1\cup N_6$ and $b\in N_1$, we may assume that $ab\notin E(G)$, then an induced path of the form $a-W_4-W_2-b$ is of length odd. Therefore, $G$ admits a harmonious cutset
by Lemma~\ref{harmonious set}. This contradiction shows that (\ref{I cap J = empty}) holds.

\medskip

Since the same conclusion holds for every choice of $X$, following from (\ref{I cap J = empty}), we see that every tail has length zero, and therefore $$\mbox{every $V$-vertex is a $Y$-vertex.}$$



Now, we show that
\begin{equation}\label{t for I J}
\mbox{there exists $t\in [7]$ such that $I\subseteq \{t-1, t, t+1\}$ and $J\subseteq \{t-3, t+3\}$.}
\end{equation}

Suppose that $5\in J$. By (\ref{IJ}) and (\ref{I cap J = empty}), we see that $\{1, 2, 3, 7\}\cap J=\emptyset$ and $5\notin I$. By Lemma~\ref{adj-Y} and (\ref{I cap J = empty}), we know that $\{4, 6\}\cap I=\emptyset$ (as otherwise $4\in I\cap J$ or $6\in I\cap J$).
By (\ref{IJ}), $\{3, 7\}\not\subseteq I$ and $\{4, 6\}\not\subseteq J$.
If $\{3,7\}\cap I=\emptyset$, (\ref{t for I J}) holds trivially from (\ref{IJ}), and then we may suppose by symmetry that $3\in I$.
If $4\notin J$, then  $t=2$ works. So, we assume that $4\in J$. By Lemma~\ref{adj-Y} and (\ref{I cap J = empty}), we know that $3\notin I$, and so $t=1$ works. This proves (\ref{t for I J}).

\medskip

By (\ref{t for I J}), we assume that $I\subseteq\{1, 2, 3\}$ and $J\subseteq\{5, 6\}$.
We claim that $N(X)$ is a harmonious cutset with $(N_2, N_1\cup P_6, N_3\cup P_5)$ being a desired partition of $N(X)$.

It is obvious that $N(X)$ is a cutset. We need to check by symmetry that
\begin{itemize}
	\item \noindent if $\{a, b\}\subseteq N_1\cup P_6$, then there exists an even induced $ab$-path with interior disjoint from $N(X)$,
	\item \noindent if $N_2\ne\emptyset$ and $\{a, b\}\subseteq N_2$, then there exists an even induced $ab$-path with interior disjoint from $N(X)$, and
	\item \noindent if $N_2\ne\emptyset$, then three sets $N_2, N_1\cup P_6$ and $N_3\cup P_5$ are pairwisely complete.
\end{itemize}

For the first one, if $\{a, b\}\subseteq N_1$, then  the statement follows from (\ref{a b even path}).
If $\{a, b\}\subseteq P_6$, we may suppose $N(a)\cap N(b)\setminus N(X)=\emptyset$, let $w_a\in N_{W_6}(a)\setminus N(b)$ and $w_b\in N_{W_6}(b)\setminus N(a)$, then we have an even $ab$-path since $\{w_a, w_b\}$ is complete to $W_5$ by Definition~\ref{def-tail-t}.
If $a\in N_1$ and $b\in P_6$, let $w_6\in N_{W_6}(b)$, then $w_6$ is complete to $W_1$, and the path $aw_6b$ is even. The second is proved in (\ref{a b even path}).
For the third one, by (\ref{Ni com to Ni+1}) and Definition~\ref{def-hepta}$(4)$, we see that $N_1, N_2$ and $N_3$ are pairwisely complete.
By (\ref{P_i com to Ni-3 and Ni+3}), we know that $P_5$ is complete to $N_1\cup N_2$ and $P_6$ is complete to $N_2\cup N_3$.
By Lemma~\ref{i,i+1 adj}, we know that $P_5$ is complete to $P_6$.
So, the three sets $N_2, N_1\cup P_6$ and $N_3\cup P_5$ are pairwisely complete.
This completes the proof of the three displayed statements above, thus $G$ admits a harmonious cutset and Lemma~\ref{remaining} holds. \qed

\end{document}